\documentclass[a4paper,12pt]{article}

\usepackage{amsmath,amssymb,amsfonts,amsxtra,makeidx,graphics,graphicx,amsthm,epstopdf,epsfig}
\usepackage[all]{xy}

\newsavebox{\ns}
\newsavebox{\dbrane}
\newsavebox{\dbshort}

\newcommand{\be}{\begin{equation}}
\newcommand{\ee}{\end{equation}}
\newcommand{\beq}{\begin{equation}}
\newcommand{\eeq}{\end{equation}}
\newcommand{\ba}{\begin{array}}
\newcommand{\ea}{\end{array}}
\newcommand{\bea}{\begin{eqnarray}}
\newcommand{\eea}{\end{eqnarray}}
\newcommand{\ben}{\begin{enumerate}}
\newcommand{\een}{\end{enumerate}}
\newcommand{\bean}{\begin{eqnarray*}}
\newcommand{\eean}{\end{eqnarray*}}
\newcommand{\eref}[1]{(\ref{#1})}

\newcommand{\BC}{\mathbb{C}}

\newcommand{\BZ}{\mathbb{Z}}
\newcommand{\comment}[1]{}

\newcommand{\CN}{{\cal N}}

\newcommand{\IP}{\mathbb{P}}
\newcommand{\IC}{\mathbb{C}}

\newcommand{\CP}{\mathbb P}

\newcommand{\setall}{\setcounter{equation}{0}
        \setcounter{theorem}{0}}

\begin{document}
\pagestyle{plain}
\setcounter{page}{1}
\newcounter{bean}
\baselineskip16pt

\begin{titlepage}

\begin{center}
\today
\hfill Imperial/TP/08/AH/06 \\
\hfill Bicocca-FT-08-13\\
\hfill MPP-2008-100 \\

\renewcommand{\thefootnote}{\fnsymbol{footnote}}

\vskip 1.5 cm
{\Large \bf Tilings, Chern-Simons Theories   and M2 Branes}
\vskip 1 cm 
{\large Amihay Hanany${}^{1}$\footnote{\tt a.hanany@imperial.ac.uk}
and Alberto Zaffaroni${}^{2}$\footnote{\tt alberto.zaffaroni@mib.infn.it}}\\
{\small
\vskip 0.8cm
${}^1${\sl Theoretical Physics Group, 
The Blackett Laboratory, \\ Imperial College London, Prince Consort Road,\\ London,  SW7 2AZ,  U.K.\\ ${}$ \\} 
\vskip .08cm
${}^2${\sl Universit\`a di Milano-Bicocca and INFN, \\ sezione di Milano-Bicocca, Piazza della Scienza, 3; \\ I-20126 Milano, Italy}
}
\end{center}

\vskip 0.5 cm

\begin{abstract}
\noindent A new infinite class of Chern-Simons theories in 2+1 dimensions 
is presented using brane tilings.
The new class reproduces all known cases so far and introduces many new models that are dual to M2 brane theories which probe a toric non-compact singular CY 4-fold.
The master space of the quiver theory is used as a tool to construct the moduli space for this class and the Hilbert Series is computed for a selected set of examples.
\end{abstract}

\end{titlepage}

\setcounter{footnote}{0}
\renewcommand{\thefootnote}{\arabic{footnote}}

\newpage
\tableofcontents

\newpage

\section{Introduction}

The $AdS/CFT$ correspondence for superconformal 2+1 dimensional gauge theories is a 
fascinating subject that has not been completely explored yet and
might reveal new surprises. It is known that M theory backgrounds 
$AdS_4\times H$, with $H$ a seven dimensional Sasaki-Einstein manifold,
preserves $\CN=2$ supersymmetry \cite{Klebanov:1998hh,Acharya:1998db,Morrison:1998cs} \footnote{In 2+1 dimensions we can also have
$\CN=1$ supersymmetry but we will not consider this case.}
and are dual to superconformal 2+1 dimensional theories. The cone over $H$ is a Calabi-Yau
four-fold and the backgrounds of interest arise as near horizon geometries
of membranes sitting at the Calabi-Yau singularity. We thus have a correspondence between the infinite number of Calabi-Yau four-folds and an infinite set
of superconformal theories. The difficult part of the story is to make
this correspondence precise and find the dual 2+1 dimensional theories.
The analogous problem in 3+1 dimensions has been solved, at least for
the class of toric Calabi-Yau singularities, using the Brane Tilings  \cite{Hanany:2005ve, Franco:2005rj, Franco:2007ii} \footnote{A proposal for 2+1 dimensions is based on crystals \cite{Lee:2006hw,Lee:2007kv}.}. 
The 2+1 dimensional case is much less understood. Previous attempts to find dual pairs have focused on Yang-Mills theories flowing in the IR to superconformal fixed  points \cite{Fabbri:1999hw,dallagata,Billo:2000zr,ohtatar}. The recent frenetic activity on M2 branes suggests the 
importance of Chern-Simons terms in this context.    

The superconformal $\CN=6$ Chern-Simons theory describing $N$ membranes
on $\IC^4/\BZ_k$ was constructed in \cite{Aharony:2008ug}. The ABJM theory is
based on a $U(N)\times U(N)$ gauge group with no kinetic term and a 
Chern-Simons term with levels $k$ and $-k$. For $k=1$ the theory
has a supersymmetry enhancement to $\CN=8$ and describes $N$ membranes on
flat space. The theory constructed in \cite{Aharony:2008ug} is the
conclusion of a long activity that followed the discovery of the BGL $\CN=8$ 
supersymmetric Chern-Simons theory and the attempts to interpret and 
generalize it \cite{BGL,tong,ghost}. Other examples of superconformal Chern-Simons theories
with supersymmetry $\CN=3,4,5$ have been constructed recently \cite{Benna:2008zy,Imamura:2008nn,Hosomichi:2008jd,Hosomichi:2008jb,Terashima:2008ba,Aharony:2008gk, Jafferis:2008qz,Fuji:2008yj}.

One interesting aspect of the ABJM theory is that it is based on a quiver
theory that in 3+1 dimensions is the conifold theory. This suggests a
relation between 3+1 dimensional quiver gauge theories and 2+1 dimensional Chern-Simons models and a possible use of 3+1 dimensional tools, for example Tilings.  
A 2+1 dimensional theory dual to an $AdS_4$ M-background should have various distinctive
features. In particular, the abelian moduli space should be a four dimensional 
Calabi-Yau cone $X$ and the non abelian moduli space should be
the symmetrized product of $N$ copies of $X$ (or a modification of it). In this
paper we will indeed show how to construct infinitely many Chern-Simons theories 
with these properties. 

In fact every consistent tiling for a 3+1 dimensional theory can be
considered as a model for a 2+1 dimensional theory with $\CN=2$ 
supersymmetry and a product of $U(N)$ gauge groups. 
When we add $\CN=2$ preserving Chern-Simons terms
and let the theory flow to the IR we generically reach some IR fixed
point. The gauge fields are massive due to the Chern-Simons terms
and are not dynamical in the IR. The moduli space of supersymmetric vacua
can be obtained by analyzing D and F terms in a similar but different fashion as in the 3+1 dimensional theory. In particular, the D terms
needs to be treated in a different way in the 2+1 dimensional Chern-Simons case.
It turns out that, in the abelian case, one particular combinations of
the $U(1)$ gauge groups should not be imposed  and only acts as 
a discrete symmetry. The abelian moduli space is then a $\IC^*$ fibration
on the Calabi-Yau threefold associated with the Tiling and it turns
out that it is a toric, non-compact, singular Calabi-Yau four-fold. The non-abelian moduli
space is generically an $N$-fold symmetric product of this space.

An important role in the analysis is played by the master space of the 3+1
dimensional model \cite{Forcella:2008bb}. This is just the $G+2$ dimensional toric variety 
obtained by solving the abelian F-term conditions, where $G$ is the number of gauge groups. By modding by the
$G-1$ complexified gauge groups we obtain the Calabi-Yau three-fold associated to the Tiling. By modding by only $G-2$ complexified gauge groups 
we obtain the four-dimensional moduli space
of the 2+1 dimensional model. In particular, every tiling gives rise to a $G-1$ dimensional family (in fact a lattice) of Calabi-Yau four-folds. In fact the direction of the $\IC^*$ action, which
is not modded out, inside the space of all $G-1$ $\IC^*$ actions depends
on the $G-1$ Chern-Simons integer parameters. The knowledge of the
master space for the 3+1 dimensional model is extremely important also to compute
the Hilbert series and the properties of the resulting four-dimensional
Calabi-Yau singularity.

The paper is organized as follows. In Section 2 we discuss the general construction and the moduli space of vacua of 2+1 dimensional Chern-Simons  theories based on brane tilings. We then proceed with examples. In Section 3 we discuss the ABJM theory from the perspective of the master space.
In Section 4,5,6 and 7 we discuss the 2+1 Chern-Simons theories associated  with tilings corresponding to the modified $\IC^2/\BZ_2\times \IC$, SPP, $\IC^3/\BZ_3$ and
$\mathbb{F}_0$ models. We end with comments and conclusions.

\section{Tilings for 2+1 Dimensional CS Theories}

Take a periodic, bipartite, two dimensional tiling of the plane, $T$, that gives rise to a consistent 3+1 dimensional theory \cite{Hanany:2005ve,Franco:2007ii}. We want to interpret it as a 2+1 dimensional theory by considering a collection of $N$ D2 branes in Type IIA instead of $N$ D3 branes in Type IIB. 
The theory has generically $\CN=2$ supersymmetry. The rules for writing down the 2+1 dimensional theory follow the rules set out for the 3+1 dimensional theory. 
As in 3+1 dimensions, every face is a $U(N)$  gauge group and
every edge is a chiral superfield transforming in a bi-fundamental representation of the two gauge groups it separates with orientation defined by the bipartite structure of the tiling. By convention one can pick an (incoming) outgoing arrow to correspond to an (anti)-fundamental representation, respectively. For each edge $E_i$ one can add an $\CN=2$ preserving Chern-Simons interaction with the following rule: add an integer CS coefficient $k_i$ and $-k_i$ to the adjacent gauge groups connected by the edge. Call $k_a$ the resulting CS coefficient for the $a$-th gauge group.

The main difference with the 3+1 dimensional case is that all gauge couplings become infinite in the IR and the standard kinetic terms for the
gauge fields can be neglected. We are left with auxiliary vector multiplets
coupled by Chern-Simons interactions. We can write the Lagrangian in $\CN=2$
superspace notations as
\beq
{\rm Tr}\left ( -\int d^4\theta \sum_{X_{ab}} X_{ab}^\dagger e^{-V_a} X_{ab} e^{V_b}  - i  \sum_a k_a\int_0^1 dt  V_a \bar D^\alpha (e^{t V_a}D_\alpha e^{-t V_a}) 
 +  \int d^2\theta  W(X_{ab}) + {\rm c.c.}\right )
\eeq
where the $V_a$ are the vector supermultiplets and  $X_{ab}$ denote chiral supermultiplets
 transforming in the fundamental representation of the gauge group $a$  and in
the anti-fundamental of the gauge group $b$. The superpotential is obtained with
the same rules of the 3+1 dimensional theory. Every vertex in the tiling contributes a term
in the superpotential given by the products of all the fields that meet at the vertex, with a positive sign for white vertices and a negative sign for black ones.

Recall that in 2+1 dimensions a vector superfield has the expansion
\beq
V= -2 i \theta\bar \theta \sigma + 2 \theta \gamma^\mu \bar\theta A_\mu + \cdot\cdot\cdot + \theta^2\bar\theta^2 D
\eeq
where we omitted the fermionic part. Compared to 3+1 dimension, there is a 
new scalar field $\sigma$. We can write the relevant terms in the Lagrangian
\beq
{\rm Tr} \left (- 4 \sum_a  k_a \sigma_a D_a + \sum_a  D_a \mu_a(X)  -\sum_{X_{ab}} (\sigma _a X_{ab} - X_{ab} \sigma_b)(\sigma _a X_{ab} - X_{ab} \sigma_b)^\dagger  -\sum_{X_{ab}} |\partial_{X_{ab}} W|^2\right )
\eeq
where $\mu_a(X)$ is the moment map for the $a$-th gauge group
\beq  
\mu_a(X)= \sum_b X_{ab} X_{ab}^\dagger - \sum_c X_{ca}^\dagger X_{ca} + [X_{aa},X_{aa}^\dagger]
\eeq
which in 3+1 dimensions is the D-term. 

By integrating out the auxiliary fields $D_a$ we see that the bosonic potential is a sum of squares. The vacua can be found
as the vanishing conditions for the scalar potential which are a set of matrix equations
\bea
\partial_{X_{ab}} W &=& 0\nonumber\\
 \mu_a(X) &=& 4 k_a \sigma_a  \nonumber\\
\sigma _a X_{ab} - X_{ab} \sigma_b &=& 0
\eea
The solutions of these equations automatically satisfy $D_a=0$ and give supersymmetric vacua. 
We see that  F-term constraints are exactly as in the 3+1 dimensional case while the D-term constraints are modified.

Let us analyze the abelian case first. The supersymmetric conditions set all $\sigma_a$ equal to a given value $\sigma$.
The remaining equations 
\beq
\mu_a(X) = 4 k_a \sigma
\eeq
look like standard D-term equations with a set of effective FI terms $\zeta_a=4k_a \sigma$. 
Since $\sum_a k_a=0$ by construction, one of these equations is redundant. Call $G$ the number of
gauge groups. We are left with $G-1$ equations.  By taking linear integer combinations
of the equations, we can set $G-2$ equations in the form
\beq
\mu_i (X) =0 \, , \qquad\qquad i=1,...G-2
\eeq
where the index $i$ identifies $G-2$ linear combinations of the gauge group, orthogonal to the direction
of the FI parameters $\zeta_a$.  
We see that we are imposing the vanishing of the D-terms for $G-2$ $U(1)$ gauge groups. 
As in 3+1 dimensions, we can dispose simultaneously 
of the D-term constraints and the corresponding $U(1)$ gauge transformations
by modding by the complexified gauge group. 

The equation for the remaining $U(1)$ gauge field looks like a D-term condition with a FI term. However, it is not adding further constraints: 
it simply determines the value of the auxiliary field $\sigma$. 
Analogously we do not need to mod out by the remaining $U(1)$ gauge group.
As explained in detail in \cite{tong,Aharony:2008ug}, the $U(1)$ is coupled to the
overall $U(1)$ gauge field by the Chern-Simons coupling and this leaves
a discrete symmetry $\BZ_k$, where $k=\gcd (\{k_a\})$. 

The moduli space of the 2+1 abelian theory it is then easy to compute.
We first solve the F-terms constraints. This gives a $G+2$ toric variety
called {\bf the master space} \cite{Forcella:2008bb}. This part of the story is  identical to the 3+1 dimensional result. See \cite{Butti:2007jv,Forcella:2008bb} for a detailed study of this concept. It is known that the master space is a $G-1$ dimensional
$\IC^*$ fibration over the threefold associated with the tiling.
The CS parameters determine a particular direction inside $(\IC^*)^{ G-1}$.
By dividing by $G-2$ $\IC^*$ actions orthogonal to the CS direction, we obtain
a 4 dimensional non-compact Calabi-Yau manifold. The moduli space is obtained by modding by the remaining discrete $\BZ_k$ symmetry. This moduli space is interpreted as the transverse space to one M2 brane in M theory which probes the four-fold.

The non abelian case is also remarkably simple. By a gauge transformation, we can diagonalize all the
$\sigma_a$. The equations $\sigma_a X_{ab}=X_{ab}\sigma_b$ tell you that generically all the fields $X_{ab}$ are diagonal \footnote{It may happen that there are other solutions of these
equations. For example, when $M$ eigenvalues of the $\sigma_a$'s coincide, a $M$ by $M$ block in the $X_{ab}$ is not necessarily diagonal and it satisfies the equations of motion of the corresponding 3+1 model with $SU(M)$ groups; this leaves a $3M+G-1$ dimensional moduli space of solutions \cite{Butti:2007jv,Forcella:2008bb}  to be compared with the $4M$ dimensional space obtained as diagonal matrices. In simple cases  these are not new solutions, but a subset of the generic ones. It is possible however that for some values of $M$ and $G$ there are new solutions. It would be interesting to check if these correspond to new branches of the moduli space, as suggested by the difference in dimension, or alternatively, to a modification of the symmetric product.}.
This reduces the problem to $N$ copies of the abelian one. The remaining discrete gauge symmetry
corresponding to the Weyl group of $SU(N)$ implies that the moduli space is generically the $N$-fold symmetric product of the abelian one. A similar statement for the mesonic moduli space of the corresponding 3+1 dimensional theory is harder to prove.  We see that the Chern-Simons theory nicely enforces in 2+1 dimensions a structure of the moduli space which is very natural from the point of view of M2 branes. It is possible, as in 3+1 dimensions, that the moduli space for some particular 
quiver contains various different branches of the moduli space. Henceforth we avoid these subtle issues and always refer to the branch corresponding to the symmetric product, or a modification of it. 

A word of caution regards the fact that we treat the Lagrangian as classical. Here we are assuming that
the theory flows in IR to a fixed point of the renormalization group which is strongly coupled.
Unfortunately, most of the 3+1 dimensional tools for studying superconformal IR fixed points
are not available in 2+1 dimensions and even this statement is difficult to check. Moreover, we are assuming that possible corrections to the K\"ahler potential are not affecting our discussion.  For an analysis of $\CN=2$ and $\CN=3$ superconformal invariance in 2+1 dimensions we refer to \cite{Gaiotto:2007qi}.

Let us now see some examples for this construction. We will encounter some models that have already appeared in the literature as orbifold of the ABJM theory and may have enhanced $\CN=3$ or $\CN=4$ supersymmetry \cite{Benna:2008zy,Imamura:2008nn,Hosomichi:2008jb,Terashima:2008ba, Jafferis:2008qz}, as well as many other new models based on 3+1 dimensional quivers. For notational convenience we will call a tiling $T$ for a 3+1 dimensional theory and a tiling $\tilde{T}_{\{k_a\}}$ for a 2+1 dimensional theory with a collection of CS terms, $\{k_a\}$, such that $\sum_a k_a = 0$. For example, a 2+1 dimensional theory resulting from the 3+1 dimensional tiling of $\mathbb{F}_0$ will be called $\tilde{\mathbb{F}}_{0\{k_1, k_2, k_3, k_4\}}$.

\section{One Hexagon, one Diagonal -- Modified Conifold} \setall

The simplest tiling contains one hexagon and the construction above is too trivial as each edge contributes zero CS to the gauge group. Therefore the simplest tiling with an interesting effect corresponds to one hexagon and one diagonal across it. There are three such diagonals and all are equivalent and lead to the same gauge theory.
The periodic tiling of one hexagon is well known to give the ${\cal N}=4$ theory in 3+1 dimensions \cite{Hanany:2005ve}.
After adding a diagonal on this hexagon \cite{Franco:2005rj}, the resulting tiling corresponds to ${\cal C}$, the conifold theory, a quiver with two gauge groups, fields $A_i,B_i,\, i=1,2$ transforming in the $(N,\bar N)$ and $(\bar N, N)$ representation of the gauge group, respectively, and interacting with the superpotential
\beq W= A_1 B_1 A_2 B_2 - A_1 B_2 A_2 B_1 \eeq

This is the $\CN=6$ theory analyzed in \cite{Aharony:2008ug}. Let us look at it using the language of Tilings and master space.

The master space for the conifold theory was computed in \cite{Forcella:2007wk} and turns out to be $\IC^4$. Such a tiling has $G=2$ gauge groups and therefore by the construction above there are no further D terms to divide by and we recover the result \cite{Aharony:2008ug} that the moduli space for the theory is $\IC^4$.
The generators are $A_1, A_2, B_1, B_2$ corresponding to the fundamental fields of the conifold theory.

The moduli space for higher $k$ is then given by the $\BZ_k$ action $1, 1, -1, -1$ on the generators, respectively.
Let us compute the Hilbert Series for this model. We will take a different approach to that of \cite{Hanany:2008qc} by picking a specific complex structure on the moduli space, which is consistent with picking a particular ${\cal N}=2$ (4 supercharges) subset out of the ${\cal N}=6$ supersymmetry of this theory in 2+1 dimensions. This approach allows for the generalization that is discussed in this paper.

The Hilbert Series can be computed easily by following the methods outlined in \cite{Hanany:2008qc} and previous papers. We use the discrete Molien invariant and find
\beq \label{conI}
g \left (t, \tilde{{\cal C}}_{\{k, -k\}} \right ) = \frac{1}{k} \sum_{j=0}^{k-1} \frac{1}{(1- \omega^j t)^2 (1-\omega^{-j} t)^2} = \frac{1+t^2+2k t^k -2k t^{k+2} -t^{2k} -t^{2k+2}}{(1-t^2)^3(1-t^k)^2},
\eeq
with $\omega^k=1$. This Hilbert Series counts the $\CN=2$ BPS states of the
abelian 2+1 dimensional theory, or, equivalently, the holomorphic functions on the 
four-fold Calabi-Yau $\IC^4/\BZ_k$.

We see here a general property of the Hilbert series we will be computing in the paper.  They will be always
palindromic, $g(1/t)=t^w g(t)$ for some weight $w$, similarly to what happens for the master space in 3+1 dimensions
\cite{Forcella:2008bb}. Due to a theorem by Stanley \cite{stanley}, the palindromic property is equivalent to the Calabi-Yau condition \footnote{We should say better  the Gorenstein condition.} and this will
give a non trivial check for all our computations.

For the case of $k=1$ the Hilbert series takes a simple form,
\beq
g \left (t, \tilde{\cal C}_{\{1,-1\}} \right ) = \frac{1}{(1-t)^4},
\eeq
from which we see that we have four free generators with weight one.
We recover the well known result that the moduli space $\BC^4$ can be thought of as a fibration of a complex line bundle over $\IP^3$, namely 4 complex variables $z_i, i=1,2,3,4$
which are subject to the identification $z_i\eqsim \lambda z_i$, for $\lambda$ some complex parameter.
This procedure will help us identify the  the complex 3 dimensional compact manifold $B$ over which the non-compact CY 4-fold is fibered over. 

For large values of $k$ the Hilbert Series behaves like
\beq
\label{conikinf}
g \left (t, \tilde{\cal C}_{\{k, -k\}} \right )  = \frac{1+t^2}{(1-t^2)^3}\left(1+O(t^k)\right ),
\eeq
which is the Hilbert series for the conifold. The large $k$ limit is equivalent
to dividing the four-dimensional moduli space by the $\IC^*$ action specified
by the charges under the remaining $U(1)$ gauge group. We see that the four-dimensional Calabi-Yau is a $\IC^*$ fibration over the three-dimensional Calabi-Yau that is associated with the tiling.

Notice that we have two different fibrations, one over a non-compact
three-fold Calabi-Yau, the conifold, and the other over a compact K\"ahler manifold, $\IP^3$. The conifold is naturally associated with the tiling, while
the compact $\IP^3$ is the relevant manifold for the Type IIA description
of the ABJM theory. When restricted to the Sasaki-Einstein seven-manifold
$H=S^7$, $\BZ_k$ is acting on the fiber of a principal $U(1)$ bundle with base
$\IP^3$. For large $k$, the length of the circle is reduced by a factor
of $k$ and, in the limit $k\rightarrow\infty$,  we can descend to 
a compactification of Type IIA on $\IP^3$ with fluxes \cite{Aharony:2008ug}.

Let us elucidate briefly on a relation between $\IP^3$ and the conifold.
In \cite{Hanany:2008qc} we wrote a partition function for $\CN=6$
chiral multiplet on $AdS_4\times S^7/\BZ_k$. The result for $k\rightarrow
\infty$
\beq \sum_{n=0}^\infty [n,0,n] t^{2 n} \eeq
where $[n,0,n]$ denotes an $SU(4)$ representation, can be interpreted
as the partition function for $\CN=6$ 
chiral multiplets in the KK compactification on $\IP^3$. 
It is well known indeed that the KK chiral multiplets for $AdS_4\times \IP^3$
fall in $[n,0,n]$ representations \cite{pope}. 
The Hilbert series (\ref{conI}) and (\ref{conikinf}) can be analogously
interpreted as the partition functions for the $\CN=2$ KK chiral multiplets
on $AdS_4\times S^7/\BZ_k$ and $AdS_4\times \IP^3$, respectively.
They differ from equations (3.15) and (3.16) of \cite{Hanany:2008qc} since here we are counting only an ${\cal N}=2$ subset of the protected operators in ${\cal N}=6$ supersymmetry. For example, out of $\dim[n,0,n] = (n+1)^2 (n+2)^2 (2n+3) / 12$ protected operators in ${\cal N}=6$ there are precisely $(n+1)^2$ operators which are holomorphic under the ${\cal N}=2$ subgroup.
We therefore sum
\beq
\sum_{n=0}^\infty (n+1)^2 t^{2n} = \frac{1+t^2}{(1-t^2)^3} ,
\eeq
and get the result computed in \eref{conikinf}.

Near $t=1$ the Hilbert series (\ref{conI}) looks like
\beq
g \left (t, \tilde{\cal C}_{\{k, -k\}} \right ) \sim \frac{1}{ k (1-t)^4},
\eeq
typical to a moduli space $\BC^4/\BZ_k$ that has a volume reduced by a factor $k$.

\section{Two Hexagons -- Modified $\BC^2/\BZ_2\times \BC$}

The theory has two gauge groups, two adjoint fields $\Phi_i$ and four chiral fields $A_i,B_i,\, i=1,2$ transforming in the $(N,\bar N)$ and $(\bar N, N)$ representation of the gauge group, respectively, and interacting with the superpotential
\beq W=  \Phi_1(A_1 B_2 - A_2 B_1) +\Phi_2 ( B_2 A_1 - B_1 A_2)  \eeq

This theory has $G=2$ gauge groups and we can modify it by adding a CS term $k$ to one of the gauge groups and $-k$ to the other.
Since the number of gauge groups is 2 the moduli space of this theory for $k=1$ is the master space of the 3+1 dimensional theory.
The largest irreducible component of the master space is computed to be ${\cal C}\times\BC$ \cite{Forcella:2007wk, Forcella:2008bb} and has a Hilbert Series
\beq
g \left (t, \widetilde{\BC^2/\BZ_2\times \BC}_{\{1,-1\}} \right ) = \frac{1+t}{(1-t)^4} .
\eeq
It is a complete intersection moduli space which is generated by 5 generators of weight 1 that are subject to 1 relation of weight 2. The generators and relations can be written in an explicit form as 4 of the bi-fundamental fields of this theory, $A_1, A_2, B_1, B_2$, subject to the quadratic relation 
\beq
\label{con}
A_1 B_2 = A_2 B_1
\eeq
generating the conifold, together with the adjoint field, $\Phi=\Phi_1= \Phi_2$, parametrizing $\BC$.
This description helps to identify the moduli space as a fibration of a line bundle over a compact manifold that is given by the relation \eref{con} of weight 2 in $\IP^4$.
Alternatively one can think of this compact manifold as a $T^2$ fibration over $\IP^1\times \IP^1$.
For higher values of $k$ we need to divide by a $\BZ_k$ discrete group that acts on the generators as one of the gauge groups: $1,1,-1,-1,0$.
The resulting Hilbert series takes the form
\beq
g \left (t, \widetilde{\BC^2/\BZ_2\times \BC}_{\{k,-k\}} \right ) = \frac{1}{k} \sum_{j=0}^{k-1} \frac{1+t}{(1- \omega^j t)^2 (1-\omega^{-j} t)^2} = \frac{1+t^2+2k t^k -2k t^{k+2} -t^{2k} -t^{2k+2}}{(1-t)(1-t^2)^2(1-t^k)^2} .
\eeq

We see that the partition function is palindromic and this indicates that the moduli space is indeed
a Calabi-Yau. For $k\rightarrow\infty$ we obtain
\beq
g \left (t, \widetilde{\BC^2/\BZ_2\times \BC}_{\{k,-k\}} \right ) = \frac{1+t^2}{(1-t)(1-t^2)^2} \left(1+O(t^k)\right ),
\eeq
which is indeed, as in the previous section, the Hilbert series of the three-fold associated with the tiling, in this case
$\IC^2/\BZ_2 \times \IC$, over which the moduli space is fibered.

\section{Two Hexagons and one Diagonal -- Modified SPP} \setall

This theory has three gauge groups, one adjoint field $\Phi$ and chiral fields, $C_1, A_1,B_1$ transforming in the $(N,\bar N,0), (0,N,\bar N)$ and $(\bar N,0, N)$ representation of the gauge group, 
respectively, and $C_2, A_2, B_2$  transforming in the complex conjugate representation,
and interacting with the superpotential
\beq W=  \Phi ( A_1 A_2  - C_2 C_1) - A_2 A_1 B_1 B_2 + C_1 C_2 B_2 B_1 \eeq

This theory has $G=3$ gauge groups and the most general CS terms given by the construction above can be denoted as $k_1, k_2-k_1, -k_2$. One of the gauge groups contains an adjoint field and by definition we can set its CS term to be the middle one, $k_2-k_1$.
Let us take a simplified model in which $k_1=k_2$.
The moduli space is computed as follows. The master space for the SPP theory is ${\cal C}\times \IC^2$, where ${\cal C}$ is the conifold \cite{Forcella:2008bb}.
It is a five dimensional complete intersection moduli space and is generated by 6 fields $A_i, B_i, C_i, i=1,2$, each of weight $t$, subject to one constraint of weight $t^2$,

\beq
A_1 A_2 = C_1 C_2.
\eeq

The number of gauge groups for the SPP theory is $G=3$ and therefore we need to divide the master space by a D term. Since we deal with the case of $k_1=k_2$ we see that the gauge group corresponding to the hexagon in the SPP tiling is a good choice.
The generators of the master space carry the following charges under the corresponding gauge group,
$1,-1, 0, 0, -1,1$. To divide by this gauge group let us construct invariants, 
\beq
M_{11} = A_1 A_2, M_{21} = C_1 A_1, M_{12}=A_2 C_2, M
_{22}=C_1 C_2,
\eeq
with the constraint
\beq
M_{12} M_{21} = M_{11} M_{22} = M_{11}^2
\eeq

The other generators of the moduli space are not affected.
The resulting moduli space is identified as $\IC^2/\BZ_2 \times \IC^2$, as already computed in \cite{Imamura:2008nn}.

The Hilbert series for this moduli space can be easily computed by introducing a weight $z$ to the charges of the gauge group we divide by and using the Molien Weyl formula,
\beq
g \left (t, \widetilde{SPP}_{\{1,0,-1\}}\right ) = \frac{1}{2\pi i} \oint_{|z|<1} \frac{dz}{z} \frac{1-t^2}{(1-t z)^2 (1-t)^2 (1-t/z)^2} = \frac{1+t^2}{(1-t)^2 (1-t^2)^2}
\eeq

The generators of this moduli space are 2 fields $B_i$ of weight 1 that generate $\BC^2$ and 3 fields $M_{ij}$ of weight 2 subject to one constraint of weight 4 that generate $\BC^2/\BZ_2$. This identifies a compact 3 dimensional manifold $B$ over which the moduli space is fibered to be the complete intersection manifold given by 1 relation of order 4 in weighted projective space, $\IP^4_{1,1,2,2,2}$. It would be interesting to see how if this is related to a Type IIA 
compactification as in \cite{Aharony:2008ug}.

For higher values of the CS term $k$ we need to divide this moduli space by $\BZ_k$ by identifying the orbifold action on the generators.
This action is identified with one of the charges of the rectangles in the SPP tiling and acts on the generators as $1, -1$ for $B_{1,2}$ and $1,0, -1$ on $M_{12} , M_{11},  M_{21}$, respectively.
The corresponding Hilbert Series can be computed using the Molien invariant
\beq
g \left (t, \widetilde{SPP}_{\{k,0,-k\}}\right ) = \frac{1}{k}\sum_{j=0}^{k-1} \frac{1+t^2}{(1- \omega^j t)(1- \omega^{-j} t) (1- \omega^j t^2) (1- \omega^{-j} t^2)} ,
\eeq
with $\omega^k=1$.

Next we turn to the more general case in which $k_1\ne k_2$ and take $\gcd(k_1,k_2)=1$. We can form a linear combination of charges that sets the D term equation to zero by taking, for example, $k_2$ copies of the first $U(1)$ and $k_1$ copies of the third. We use Table \ref{globalconifold} to compute the charges.

\begin{table}[htdp]
\caption{Global charges for the basic fields of the quiver gauge theory on the D brane probing the SPP singularity.}
\begin{center}
\begin{tabular}{|c||c|c|c|c|c||c|}
\hline
 & $U(1)_1$ & $U(1)_2$ & $U(1)_3$ & $k_2 U(1)_1 + k_1U(1)_3$ \\ \hline \hline
$A_1$ & $0$ & $1$ & $-1$ & $-k_1$ \\ \hline
$A_2$ & $0$ & $-1$ & 1 & $k_1$ \\ \hline
$B_1$ & $-1$ & 0 & $1$ & $k_1-k_2$ \\ \hline
$B_2$ & 1 & 0 & $-1$ & $k_2-k_1$  \\ \hline
$C_1$ & 1 & $-1$ & $0$ & $k_2$  \\ \hline
$C_2$ & $-1$ & $1$ & $0$ & $-k_2$ \\ \hline
\end{tabular}
\end{center}
\label{globalconifold}
\end{table}
This leads to the Hilbert Series
\bea
&&g \left (t, \widetilde{SPP}_{\{k_1,k_2-k_1,-k_2\}}\right ) = \\ &=& \frac{1}{2\pi i} \oint_{|z|<1} \frac{dz}{z} \frac{1-t^2}{(1-t z^{-k_1}) (1-t z^{k_1})(1-t z^{k_1-k_2})(1-t z^{k_2-k_1})(1-t z^{k_2}) (1-t z^{-k_2})} \nonumber
\eea
One for example can take the case of $k_1=1, k_2=2$ and get 
\beq
g \left (t, \widetilde{SPP}_{\{1,1,-2\}}\right ) = \frac{1+2 t^2 +4 t^3+2 t^4 + t^6}{(1-t ^2)^2(1-t ^3)^2} ,
\eeq
giving rise to a non-compact CY manifold of dimension 4 which is generated by 4 operators of weight 2 and 6 of weight 3 that are subject to 1 relation at weight 4, 8 relations at weight 5 and 11 relations at weight 6. We see that the Hilbert
series is palindromic, as expected. In the case $\gcd(k_1,k_2)\ne 1$ we will need to mod by an extra discrete symmetry,
which can be done as in the previous case by using the discrete Molien invariant.

\section{The simplest chiral model -- Modified $\IC^3/\BZ_3$}

All model studied so far and most of the models studied in the literature are non-chiral where chirality is understood to be in the 3+1 dimensional sense. This means that in the quiver description of the model, the non-chiral theory has equal number of arrows going to and from each pair of nodes, while in the chiral theory it is not.

Consider now a chiral model with three hexagons. There are three groups and three sets of chiral fields $U_i,V_i,W_i, \, i=1,2,3$ transforming in the $(N,\bar N,0), (0,N,\bar N)$ and $(\bar N,0, N)$ representation of the gauge group, 
respectively, and interacting with the superpotential

\beq W= \epsilon_{ijk} U_i V_j W_k  \eeq 

We can put different CS terms
$k_1,k_2,-k_1-k_2$. The master space, computed in \cite{Forcella:2008bb}
is particularly simple, given by the variety $\IC^6/\{1,1,1,-1,-1,-1\}$.
This has dimension five and it is a $(\IC^*)^2$ fibration over $\IC^3/\BZ_3$.
The linear combination $k_2 U(1)_1-k_1 U(1)_2$ of gauge groups has a vanishing 
D-term. The corresponding action on $\IC^6$ is $\{k_2+k_1,-k_2,-k_1,0,0,0\}$.
Let us first take $\gcd(k_1,k_2)=1$. We can write the Hilbert series for $(\widetilde {\IC^3/\BZ_3})_{k_1,k_2}$ as
\beq
g \left ( t,\widetilde{\IC^3/\BZ_3}_{\{k_1,k_2, -k_1-k_2\}} \right ) = \frac{1}{(2\pi i)^2} \oint  \frac{dz}{z}\frac{dw}{w} \frac{1}{(1-  z w^{k_2+k_1})(1- z w^{-k_2})(1- z w^{-k_1})(1-t/z)^3} .
\eeq

More generally, for $k_1=k m_1,k_2=k m_2$ with $\gcd(k_1,k_2)=k$, the linear combination $m_2 U(1)_1-m_1 U(1)_2$ of gauge groups has a vanishing D-term and we have
to mod by a $\BZ_k$ subgroup of the orthogonal gauge group $-U(1)_3$

\bea
&&g \left ( t,\widetilde{\IC^3/\BZ_3}_{\{k_1,k_2, -k_1-k_2\}} \right ) = \\ &=& \frac{1}{k}\sum_{j=0}^{k-1}\frac{1}{(2\pi i)^2} \oint  \frac{dz}{z}\frac{dw}{w} \frac{1}{(1-  z w^{m_2+m_1})(1- z \omega^{-j} w^{-m_2})(1- z \omega^j w^{-m_1})(1-t/z)^3} , \nonumber
\eea
where $\omega^k=1$.

\section{Another chiral model -- Modified $\mathbb{F}_0$}

The gauge theory is the usual quiver theory for $\mathbb{F}_0$ and for simplicity we will study phase $I$ of this theory \cite{Forcella:2008bb}. 
We have four groups and four sets of chiral fields $A_i,B_i,C_i,D_i \, i=1,2$ transforming in the $(N,\bar N,0,0), (0,N,\bar N,0),(0,0,N,\bar N) $ and $(\bar N,0,0, N)$ representation of the gauge group, 
respectively. The superpotential is
\beq W =  \epsilon_{ij} \epsilon_{pq} A_i B_p C_j D_q \eeq

The master space of $\mathbb{F}_0$ is a product of two conifolds, ${\cal C}\times {\cal C}$ \cite{Forcella:2008bb}, and therefore it is a complete intersection generated by 8 generators $A_i, B_i, C_i, D_i$ of weight $t$ subject to two relations of weight 2,
\beq
A_1 C_2 = A_2 C_1, \qquad B_1 D_2 = B_2 D_1 .
\eeq

The corresponding Hilbert series for the master space is
\beq
H(t, {\cal C}\times {\cal C} ) = \frac{(1-t^2)^2}{(1-t)^8} .
\eeq

Let us first consider the model where we add CS terms $k$ and $-k$ for two gauge groups only, say the first and fourth. To get the moduli space of the 2+1 dimensional theory we need to divide the master space by two D term conditions, coming from the gauge groups with CS terms equal to 0.
The charges under these two gauge groups are for $A, B, C, D$ $-1, 1, 0, 0$ and $0, -1, 1, 0$, respectively.
To compute the generators of the resulting moduli space it is possible to follow the procedure outlined above for the case of the SPP theory but it is more systematic to use the Hilbert Series. So let us first compute the Hilbert Series for this theory. Using the Molien-Weyl integral we have
\bea
g \left (t, \tilde{\mathbb{F}}_{0 \{1,0,0,-1\}} \right ) &=& \frac{1}{(2\pi i)^2} \oint_{|z_1|<1} \frac{dz_1}{z_1} \oint_{|z_2|<1} \frac{dz_2}{z_2} \frac{(1-t^2 z_2/z_1)(1-t^2 z_1/z_2)}{(1-t/z_1)^2 (1-t z_1 / z_2)^2 (1-t z_2)^2 (1-t)^2 } \nonumber \\
&=& \frac{1+t+t^2+4t^3+t^4+t^5+t^6}{(1-t) (1-t^3)^3} ,
\eea
indicating that the moduli space is indeed 4 dimensional, as expected. The palindromic Hilbert series also implies that this is a CY manifold. By construction it is further toric.
To get more information on this moduli space we take the Plethystic Logarithm which gives information on the generators and relations,
\beq
PL \left [ g \left (t, \tilde{\mathbb{F}}_{0 \{1,0,0,-1\}} \right )  \right ] = 2t+6t^3-3t^4 -6t^6+O(t^7)
\eeq
This means that there are 2 generators of weight 1 which are identified with $D_i$ and 6 generators of weight 3 which are identified with the gauge invariant combinations $A_i B_j C_l$, with $i,l$ symmetrized due to the F term conditions, giving all together 6 generators.
There are 3 relations at weight 4 and 6 relations at weight 6. This is the moduli space of the modified $(\tilde{\mathbb{F}}_0)_{\{1,0,0,-1\}}$ theory. We can further view it as a fibration over a 3 dimensional complex base $B$ given by the non-complete intersection of 3 relations of weight 4 and 6 relations of weight 6 in the weighted projective space, $\CP^7_{1,1,3,3,3,3,3,3}$ .

Next we turn to the case $k>1$. This is a $\BZ_k$ orbifold of the previous theory. The action on the generators is 1 on $D$'s and $-1$ on the $ABC$'s. 
Since this moduli space is not a complete intersection it contains infinitely many syzygies and it is hard or impractical to figure out the action of $\BZ_k$ on each of them. Instead we will refer to the action of $\BZ_k$ on the basic fields of the theory as they appear in the tiling. We recall that $D$ has charge $1$ and $C$ has charge $-1$. This allows to write the combination of Molien-Weyl integral and Molien invariant,

\bea
g \left (t, \tilde{\mathbb{F}}_{0 \{k,0,0,-k\}} \right ) &=& \frac{1}{(2\pi i)^2 k} \sum_{j=0}^{k-1} \oint_{|z_1|<1} \frac{dz_1}{z_1} \oint_{|z_2|<1} \frac{dz_2}{z_2} \frac{(1- \omega^{-j} t^2 z_2/z_1)(1- \omega^j t^2 z_1/z_2)}{(1-t/z_1)^2 (1-t z_1 / z_2)^2 (1- \omega^{-j} t z_2)^2 (1- \omega^j t )^2 } \nonumber \\
&=& \frac{1}{k} \sum_{j=0}^{k-1} \frac{1+ 3 \omega^{-j} t^3 - 3 t^4 - \omega^{-j} t^7}{(1- \omega^j t)^2 (1- \omega^{-j} t^3)^3} = \frac{1+6t^4+t^8+(k+1)t^k+\ldots+t^{9k+8}}{(1-t^4)^3(1-t^{3k})^3} .
\eea
The last sum is doable but is too long to report here.
The numerator is indeed verified to be a palindromic polynomial of order $9k+8$, indicating that the moduli space is CY for any $k$, as expected.
By taking the PL of this expression we find 9 generators at weight 4 corresponding to the 9 basic mesonic operators of $\mathbb{F}_0$. These operators satisfy 20 relations, again as expected from $\mathbb{F}_0$. The new feature comes from $k+1$ generators at weight $k$ that obviously are a consequence of the CS term.
The Hilbert series has a behavior at large $k$ that goes like
\beq
g \left (t, \tilde{\mathbb{F}}_{0 \{k,0,0,-k\}} \right ) \sim \frac{1+6t^4+t^8}{(1-t^4)^3} (1+ O(t^k))
\eeq
This is indeed the Hilbert series for the three dimensional Calabi-Yau cone over $\mathbb{F}_0$ \cite{BFHH}. We see again that in the $k\rightarrow\infty$ limit we recover the Hilbert series of the three-fold associated with the tiling, over which the four-fold is a $\IC^*$ fibration.

As a second example, we can consider the theory with CS parameters $k,-k,k,-k$. The two gauge groups $U(1)_1+U(1)_2$ and $U(1)_1+U(1)_4$ acting on the fields $A, B, C, D$ with charges $0, 1, 0, -1$ and 
$1, 0, -1, 0$ have zero D-terms and can be modded out. For $k=1$ we obtain the remarkably simple result
\beq
g \left (t, \tilde{\mathbb{F}}_{0 \{1,-1,1,-1\}} \right ) = \frac{1}{(2\pi i)^2} \oint_{|z_i|<1} \frac{dz_1}{z_1}  \frac{dz_2}{z_2} \frac{(1-t^2)^2}{(1-t z_2)^2 (1-t z_1)^2 (1-t/ z_2)^2 (1-t/ z_1)^2 } = \frac{(1+t^2)^2}{(1-t^4)^4} ,
\eeq
corresponding to the variety $\IC^2 /\BZ_2\times \IC^2/\BZ_2$. This model is presumably one of the chiral orbifolds
considered in \cite{Benna:2008zy}. For $k>1$ we mod by the discrete action of the gauge group $U(1)_3$
\beq
g \left (t, \tilde{\mathbb{F}}_{0 \{k,-k,k,-k\}} \right ) = \frac{1}{k} \sum_{j=0}^{k-1}\frac{1}{(2\pi i)^2} \oint_{|z_i|<1} \frac{dz_1}{z_1} \frac{dz_2}{z_2} \frac{(1-t^2)^2}{(1-t z_2)^2 (1-t z_1\omega^{-j})^2 (1-t \omega^j/ z_2)^2 (1-t/ z_1)^2 } \eeq
with $\omega^k=1$.

We could similarly analyze the case with generic CS parameters $k_1,k_2,k_3,-k_1-k_2-k_3$ giving a three integer
family of four-fold Calabi-Yau singularities. 

\section{Dimers and Orientifolds}

For this case we will be brief, leaving a detailed discussion for future work. One can construct CS theories in 2+1 dimensions by following the same construction in this paper and apply it to the tilings which were introduced in \cite{Franco:2007ii}. The simplest model corresponds to an $Sp\times SO$ model which was discussed in the literature in various papers \cite{Hanany:2008qc,Hosomichi:2008jb,Aharony:2008gk}, a prototypical case leading to an orbifold $\IC^4/D_k$ moduli space. A particularly interesting aspect about this class of models is that they lead to non-toric CY 4-folds and therefore are a very interesting tool in studying such backgrounds beyond the familiar toric tools. Other interesting models
are obtained as non-supersymmetric orientifolds \cite{Armoni:2008kr}.

\section{Comments and Conclusions}

In this paper we demonstrate how to construct infinitely many 2+1 dimensional product $U(N)$ Chern-Simons theories with moduli space which is generically the $N$-fold symmetric product of toric
four-fold Calabi-Yau singularities. The natural expectation is that these theories
are dual to the M theory background $AdS_4\times H$, where $H$ is the 7 real dimensional Sasaki-Einstein base of the Calabi-Yau. 

We explicitly computed the Hilbert series for the abelian theory. Under the assumption that the moduli space is a symmetric product we can compute the Hilbert series - the partition function for chiral operators - for all values of $N$ using the Plethystic Exponential \cite{BFHH,Hanany:2006uc,feng}. This computation is extremely simple, similar to the computation of the partition function for the mesonic moduli space in 3+1 dimensions and does not involve all the complications due the existence of baryonic directions \cite{Butti:2006au,Forcella:2007wk}.

More generally, our construction opens the way for many interesting investigations.

Of course the most important point would be  to prove that the CS theory is really dual to the $AdS_4$ background; this might be difficult because many efficient 3+1 dimensional tools to study superconformal theories are not available in 2+1 dimensions at the moment of the writing. We could start by trying to understand better the existence and the properties of the IR fixed points corresponding to  Chern-Simons theories. Similarly to the generic 3+1 dimensional quiver, these theories will be strongly coupled and with a spectrum of conformal dimensions that is not the canonical one.
In the 3+1 dimensional case, we can explicitly compare
the results of a-maximization  \cite{Intriligator:2003jj,Barnes:2005bm} with the spectrum of dimensions computed
from volumes of cycles in $H$  and gain confidence in our identification
of the quiver gauge theory as the dual of $AdS_4\times H$ \cite{Bertolini:2004xf,Benvenuti:2004dy,Benvenuti:2005ja,Franco:2005sm,Butti:2005sw}.
This comparison can be done for a generic toric three-fold \cite{Butti:2005vn} and it
is a highly non-trivial check of the correspondence between Tilings and 
toric Calabi-Yau singularities. In 2+1 dimensions, the familiar tools for studying the IR fixed point and 
the exact R-charges and dimensions, as for example a-maximization,
are not available. We can still compute volumes
of cycles in toric Calabi-Yau singularities \cite{ Martelli:2005tp,Martelli:2006yb} 
and extract from this the dimension
of a subset of fields. It would be quite interesting to find a surrogate
of a-maximization to compare with. Less ambitiously, it would be interesting
to understand if and how we can parametrize R-charges using
the toric data of the Calabi-Yau singularity, as it happens in 3+1 dimensions \cite{Benvenuti:2005ja,Franco:2005sm,Butti:2005sw,Butti:2005ps}.  

Another interesting point regards the Type IIA description of the 
supergravity background. The discrete group $\BZ_k$, with 
$k= \gcd (\{k_a\})$, is acting on the seven-manifold $H$. In the
ABJM case, it acts on the fiber of a $U(1)$ bundle on $\IP^3$ 
and it reduces the length of the circle by a factor of $k$. For large $k$,     
it is better to descend to Type IIA compactified on $\IP^3$ with fluxes.
It would be nice to understand this construction for a generic four-fold,
to see whether we can always descend in Type IIA preserving $\CN=2$
supersymmetry and identify the compact three-manifold $B$ that appears in the 
supergravity solution. Notice, in particular, that we considered
two different fibrations of the four-fold: one over the non-compact Calabi-Yau
associated with the Tiling, and another over a compact K\"ahler three manifold.
It would be quite interesting to understand the relation between these
different three dimensional algebraic varieties. By consistency,
we can expect that the partition function for chiral scalar KK modes on $B$
(eigenfunctions of the Laplacian on $B$) coincides with the partition
function for holomorphic functions on the Calabi-Yau three-fold.   

Finally, the most ambitious goal would be to find a complete classification
of the 2+1 dimensional theories dual to toric (and non toric) Calabi-Yau singularities. In particular, in this paper we produce a direct algorithm to generate Calabi-Yau singularities for a given
Chern-Simons theory. A natural question is if all toric Calabi-Yau singularities  
arise in this way and how to find the inverse algorithm, that in 3+1 dimensions is developed in \cite{Feng:2000mi, Hanany:2005ss, Gulotta:2008ef}. See \cite{Agarwal:2008yb} for further details. This algorithm takes the toric data of a given Calabi-Yau singularity and generates the 3+1 dimensional theory. Similarly we would like to take the toric data of a 4-fold CY singularity and generate the 2+1 dimensional theory. 

We plan to investigate all these issues in the future with 
many other interesting topics that are missing in the list.

\vskip 1 truecm

\noindent {\bf Note added:} While finishing this work, a paper \cite{Martelli:2008si} appeared in the arXiv, which has 
some overlap with our results. Comments on the moduli space of $\CN=2$ CS theories also appeared in \cite{Jafferis:2008qz}.

\section*{Acknowledgments}
We would like to thank A.~Tomasiello for useful discussions.
A.~Z.~ is supported in part by INFN  and by the European Community's Human Potential Program MRTN-CT-2004-005104.
A.~H.~ would like to thank the Max Planck Institute for theoretical physics and the Ludwig Maximilian University in Munich for their kind hospitality during the writing of this paper.

\appendix

\end{document}